\documentclass[11pt]{article}
\pdfoutput=1
\usepackage{jcappub}
\usepackage{lineno}
\usepackage{csquotes}
\usepackage{comment}
\usepackage{booktabs}
\usepackage{longtable}
\usepackage{pdflscape}
\usepackage{array}
\usepackage{tabularx}
\usepackage{helvet}
\usepackage{units}
\usepackage{wrapfig}
\usepackage{indentfirst}
\usepackage{xstring}
\usepackage{amsmath}
\usepackage[font=scriptsize,labelsep=period]{caption}
\usepackage{url}
\usepackage{booktabs}
\usepackage{graphicx}
\usepackage{afterpage}
\usepackage{float}
\usepackage{indentfirst}
\usepackage{amsmath}
\usepackage{amssymb}
\usepackage{amsthm}
\usepackage{multicol}
\usepackage{multirow}
\usepackage{fancyhdr}
\usepackage{array}
\usepackage{geometry}
\usepackage{anysize}
\usepackage{lmodern}
\usepackage{ulem}
\usepackage{verbatim}

\title{Modeling the probability distribution for cosmological analysis with photometrically classified samples}

\author[a]{Marcos P. Freaza}
\author[a,b]{Ribamar R. R. Reis}

\affiliation[a]{Observatório do Valongo, Universidade Federal do Rio de Janeiro\\
Ladeira Pedro Antônio, 43, Centro,  Rio de Janeiro,
Brasil}
\affiliation[b]{Instituto de Física, Universidade Federal do Rio de Janeiro,\\
Av. Athos da Silveira Ramos, 149 - Cidade Universitária, Rio de Janeiro, Brasil}
\emailAdd{freaza@ov.ufrj.br, ribamar@if.ufrj.br}

\abstract{
In this work we investigated methods for the accurate and efficient incorporation of photometrically classified supernovae into cosmological analyses, and to assess the impact of the additional uncertainty associated with this procedure on the ability of Type Ia supernovae (SNeIa) tests to place constraints on cosmological models. We proposed a simplified likelihood, in which the contamination is described as a redshift dependent change in the mean of the usually assumed Gaussian distribution, and we tested this hypothesis against the usual two-component approach, based on the BEAMS framework. Using the latest version of the DES supernova sample, dubbed DES-Dovekie, we compared the results when using type probabilities from different classifiers, such as SNIRF and SCONE, and applying different cuts on these probabilities. We show that the new model is strongly favored by the Bayes factor, when compared with the current one, for all configurations, allowing an improvement on the constraining power of photometric supernova data.}

\begin{document}

\maketitle

\section{Introduction}

Type Ia supernovae (SNeIa) has been a key observable to study cosmic expansion for decades and provided the first evidence for acceleration \cite{riess1998,perlmutter1999}. Obtaining cosmological constraints from these events relies on three key steps, classification, light curve fitting and statistical analysis. A number of the algorithms were developed throughout the years in order to standardize supernovae \cite{phillips1993,hamuy1996,phillips1999,jha2007,burns2011,rose2020,nascimento2024} but, in the last decade, the widely used one is SALT (Supernova Adaptive Lightcurve Template) and its variations \cite{guy2005,Guy2007,guy2010,betoule2014,kenworthy2021,taylor2023}. The light curve fitting processes were designed to be applied on SNeIa only and, for a long time, various spectroscopic confirmed samples were made available for the community \cite{astier2006,Wood-Vasey_2007,kowalski2008,amanullah2010,guy2010,betoule2014,Sako2018,hoyt2026}. 

However, with increasingly large photometric surveys, the number of candidates is simply too big to allow spectroscopic followup of a significant fraction, motivating the use of photometrically classified events. This necessity stimulated the emergence of various photometric classifiers \cite{poznanski2007,connolly2009,richards2012,karpenka2013,ishida2013,moller2016,kimura2017,revsbech2018,moss2018,villar2020,santos2020,moller2020,dobryakov2021} and redshift estimators optimized for SNeIa \cite{kessler2010b,wang2015,qu2023, oliveira2023,lee2024}. This also introduced a new challenge, how to treat the contamination (events erroneously classified as SNeIa) within the statistical analyses. 

In the last years, the majority of real and simulated data has been analyzed through roughly the same framework, introduced in \cite{Hlozek2012}, and based in previous contributions such as BEAMS (Bayesian Estimation Applied to Multiple Species) \cite{Kunz_2007} and later improved with BBC (BEAMS with Bias Corrections) \cite{kessler2017}. In this work, we use propose a simplified statistical model, changing the likelihood for the photometrically classified supernova samples, but using the same inputs, namely, the distance modulus estimations from SALT3 and BBC, in order to make the analysis less sensitive to contamination in those samples.

\section{The model for the likelihood}

The model presented in \cite{Hlozek2012}, based in the BEAMS  \cite{Kunz_2007} framework, defines the probability distribution for one supernova as a combination distributions for Ia and non-Ia:

\begin{equation}
P(\mu_i \mid \boldsymbol{\theta}) =
P(\mu_i \mid \boldsymbol{\theta}, \mathrm{Ia})\, P_i
+
P(\mu_i \mid \boldsymbol{\theta}, \mathrm{nIa})\, (1 - P_i),
\label{snpdf}
\end{equation}
where $P_i$ is the probability for the ith supernova being Ia, $\mu_i$ is the distance modulus estimation for this event, $\boldsymbol{\theta}$ is the set of free parameters that specify the theoretical model, $P(d_i \mid \boldsymbol{\theta}, \mathrm{Ia})$ is the distribution for the distance, given the event is a SNIa, and $P(d_i \mid \boldsymbol{\theta}, \mathrm{nIa})$ is the distribution for the distance, given the event is NOT a SNIa.

As usual, SNIa are modeled assuming the distance moduli are distributed according a Gaussian distribution with mean given by the theoretical model,
\begin{equation}
P(\mu_i \mid \boldsymbol{\theta}, \mathrm{Ia}) =
\mathcal{N}
\left(
\mu_{th}(z_i, \boldsymbol{\theta}),
\sigma_{\mathrm{tot},i}^2
\right).
\label{Iapdf}
\end{equation}
The distance modulus is given by
\begin{equation}
\mu_{th}(z, \boldsymbol{\theta}) = 5 \log_{10}\left[\frac{D_L(z, \boldsymbol{\theta})}{\mathrm{Mpc}}\right] + 25,
\label{muIa}
\end{equation}
with the luminosity distance given by
\begin{equation}
D_{\rm L}(z,\boldsymbol{\theta}) = \left\{
\begin{array}{ll}
(1+z)\dfrac{c}{H_0}\,\frac{1}{\sqrt{\Omega_{k0}}}\,\sinh\left[\sqrt{\Omega_{k0}}\,D_{\rm C}(z,\boldsymbol{\theta})H_0/c\right] & {\rm for}~\Omega_{k0}>0 \\
(1+z)D_{\rm C}(z,\boldsymbol{\theta}) & {\rm for}~\Omega_{k0}=0 \\
(1+z)\dfrac{c}{H_0}\,\frac{1}{\sqrt{|\Omega_{k0}|}}\,\sin\left[\sqrt{|\Omega_{k0}|}\,D_{\rm C}(z,\boldsymbol{\theta})H_0/c\right] & {\rm for}~\Omega_{k0}<0
\label{dl}
\end{array}
\right. ,
\end{equation}
where
\begin{equation}
D_C(z,\boldsymbol{\theta}) = \,c \int_0^z \frac{dz'}{H(z',\boldsymbol{\theta})}
\end{equation}
is the comoving distance, and the Hubble parameter is
\begin{equation}
H(z) = H_0 \sqrt{
\Omega_{m0} (1+z)^3 + \Omega_{k0} (1+z)^2 + \Omega_{\Lambda 0}
},
\label{hubble}
\end{equation}
with $\Omega_{k0} = 1 - \Omega_{m0} - \Omega_{\Lambda 0}$.

The total variance is given by
\begin{equation}
\sigma_{\mathrm{tot},i}^2 =
\sigma_{\mu,i}^2
+
\sigma_{\mu,z,i}^2
+
\sigma_{\mathrm{int}}^2\,,
\label{Iavar}
\end{equation}
where $\sigma_{\mu,i}$ denotes the uncertainty from the light curve fit, $\sigma_{\mu,z,i}$ is the additional uncertainty in the distance due to redshift error, and $\sigma_{\mathrm{int}}$ is a potential residual intrinsic dispersion, which is considered a free parameter. Here, for simplicity, we are considering only statistical errors and neglecting correlation between different events, as a first approximation.

The uncertainty due to redshift is usually modeled as
\begin{equation}
\sigma_{z,i}^2 = 
\left[
\frac{5}{\ln 10}
\frac{1+z_i}{z_i \left(1+\frac{z_i}{2}\right)}
\right]^2
\left(\sigma_{z,i}^2 + 0.001^2\right),
\end{equation}
where $\sigma_{z,i}$ is the redshift error and the second term accounts for peculiar velocities, which we assume to be of 300 km/s.

The Non-Ia population is also described by a Gaussian distribution, but with different mean and, possibly, variance,
\begin{equation}
P(\mu_i \mid \boldsymbol{\theta}, \mathrm{nIa}) =
\mathcal{N}
\left(
\mu_{th}(z_i, \boldsymbol{\theta}) + \Upsilon(z_i,\boldsymbol{\phi}),
s_{\mathrm{tot}}^2
\right),
\label{nonIapdf}
\end{equation}
where $\Upsilon(z_i,\boldsymbol{\phi})$ denotes a deviation from the cosmological model, as a function of redshift and $s_{\mathrm{tot}}$ is the total variance associated to this population. The set of parameters $\boldsymbol{\phi}$ specifying the deviation are treated as nuisance and the final cosmological results are obtained after marginalization over them.

Therefore, the likelihood for a photometric supernova dataset, which will dub ``Hlozek model'', is 
\begin{align}
L(\boldsymbol{\theta},\boldsymbol{\phi}) & = 
\prod_{i=1}^{N}
\left[
P_i \,
\frac{1}{\sqrt{2\pi}\,\sigma_{\mathrm{tot},i}}
\exp\left(
-\frac{(\mu_i - \mu_{th}(z_i,\boldsymbol{\theta}))^2}
{2\sigma_{\mathrm{tot},i}^2}
\right) \right. \\
& \left.+
(1-P_i) \,
\frac{1}{\sqrt{2\pi}\,s_{\mathrm{tot},i}}
\exp\left(
-\frac{(\mu_i - \eta(z_i,\boldsymbol{\theta},\boldsymbol{\phi}))^2}
{2s_{\mathrm{tot},i}^2}
\right)
\right],
\label{hlozekmod}
\end{align}
where 
\begin{equation}
    \eta(z_i,\boldsymbol{\theta},\boldsymbol{\phi})=\mu_{th}(z_i, \boldsymbol{\theta}) + \Upsilon(z_i,\boldsymbol{\phi}).
    \label{nonIamean}
\end{equation}

It is worth to note that, in the original work \cite{Hlozek2012}, it is reported that the perform better without the theoretical distance modulus in the mean for the Non-Ia distribution, being $\eta(z_i,\boldsymbol{\theta},\boldsymbol{\phi})=\Upsilon(z_i,\boldsymbol{\phi})$. Here, we keep the term for completeness.
In this work, we investigate two functions for the non-Ia mean, a 3rd order polynomial
%\begin{align}
%    \Upsilon_2(z) & = a + b z + c z^2, \\
%    \Upsilon_2(z) & = a + b z + c z^2 + d z^3.
%\end{align}
\begin{equation}
    \Upsilon_2(z)  = a_0 + a_1 z + a_2 z^2 + a_3 z^3.
\end{equation}

As an alternative, we propose that the probability distribution of one supernova is given by only one Gaussian, whose mean is modified according its probability of being Ia. The basic hypothesis is that, after quality cuts, the contamination is expected to be small and, in this case, it may be possible to neglect changes in the form of the distribution. The likelihood for our model, dubbed Gaussian with Modified Mean (GMM) is written as
\begin{equation}
L_{\mathrm{GMM}}(\boldsymbol{\theta},\boldsymbol{\phi}) :=
\prod_{i=1}^{N}
\left[
\frac{1}
{\sqrt{2\pi}\sigma_{\mathrm{tot},i}}
\exp\left(
-\frac{\chi_i^2(\boldsymbol{\theta},\boldsymbol{\phi})}
{2 \sigma_{\mathrm{tot},i}^2}
\right)
\right],
\label{GMMmod}
\end{equation}
where $\sigma_{\mathrm{tot},i}$ is given by Eq. (\ref{Iavar}), $P_i$ is probability of being Ia and $\sigma_{ad}$ is an additional dispersion due to the contamination from non-Ia events, and
\begin{equation}
\chi_i^2 :=
\left\{\mu_i -
\mu_{th}(z_i,\boldsymbol{\theta})
-(1-P_i)\,f(z_i,\boldsymbol{\phi})
\right\}^2.
\label{GMMmean}
\end{equation}
The function $f(z)$ is the deviation from the model for non-Ia events. Here, we also propose a 3rd order polynomial
\begin{equation}
f(z,\boldsymbol{\phi}) = a_1 z + a_2 z^2 + a_3 z^3,
\label{poli3semmodulo}
\end{equation}
where the coefficients $\boldsymbol{\phi}=(a_1,a_2,a_3)$ are free parameters, to be fitted with the cosmological parameters and the additional dispersion $\sigma_{int}$. We dropped the constant parameter in the polynomial after verifying that the model performs better without it.

We assume standard $\Lambda$CDM throughout the work and focus on the differences between the approaches for this fixed family of models. Therefore, the free parameters are ($\Omega_{m0}$, $\Omega_{\Lambda 0}$, $a_0$, $a_1$, $a_2$, $a_3$, $\sigma_{\text{int}})$. We also consider, for GMM, a simultaneous fit with SALT3 nuisance parameters $(\alpha,\beta,M)$.

In order to compare the results for the different models, we chose to use the Bayes factor
\begin{align}
    \mathcal{B}_{ij} & = \dfrac{Z_i}{Z_j}, \\
    \ln \mathcal{B}_{ij} & = \ln Z_i - \ln Z_j,
\end{align}
where $Z_i$ is the Bayesian evidence for one model,
\begin{equation}
    Z_i = \int L_i(\boldsymbol{\theta},\boldsymbol{\phi})\Pi (\boldsymbol{\theta},\boldsymbol{\phi}) d\boldsymbol{\theta} d\boldsymbol{\phi},
\end{equation}
where $\Pi (\boldsymbol{\theta},\boldsymbol{\phi})$ is the prior, which we assume to be uniform for all parameters, and we use the usual Jeffreys scale \cite{Jeffreys:1939xee,robert2008} to classify the models, as show in table \ref{jeffreyscale}.

\begin{table}
\begin{center}
\caption{The logarithmic Jeffreys scale for the Bayes factor.} \label{jeffreyscale}
\begin{tabular}{cc}
  \toprule
  \hspace{5pt} $\ln\mathcal{B}_{ij}$ range \hspace{5pt} & \hspace{5pt} Evidence \hspace{5pt}\\
  \midrule
  \hspace{5pt} $[0,1.1]$ \hspace{5pt} & \hspace{5pt} Weak       \\
  \hspace{5pt} $[1.1,3]$   \hspace{5pt} &\hspace{5pt} Definite   \\
  \hspace{5pt} $[3,5]$   \hspace{5pt} & \hspace{5pt} Strong     \\
  \hspace{5pt} $[5,\infty)$   \hspace{5pt} & \hspace{5pt} Very Strong\\
  \bottomrule
\end{tabular}
\end{center}
\end{table}

\section{Results}
\begin{comment}

A metodologia descrita foi aplicada a diferentes subconjuntos da amostra, definidos por cortes mínimos na probabilidade \(P_i\) de classificação como supernova do tipo Ia. Os parâmetros cosmológicos estimados são apresentados a seguir:

\begin{itemize}
\item Para supernovas com \(P_i \geq 0.9\):
\[
\Omega_m = 0.23 \pm 0.10, \qquad
\Omega_\Lambda = 0.52 \pm 0.10.
\]

\item Para supernovas com \(P_i \geq 0.8\):
\[
\Omega_m = 0.234 \pm 0.097, \qquad
\Omega_\Lambda = 0.51 \pm 0.18.
\]

\item Para supernovas com \(P_i \geq 0.5\):
\[
\Omega_m = 0.246 \pm 0.091, \qquad
\Omega_\Lambda = 0.52 \pm 0.17.
\]

\item Considerando a amostra completa:
\[
\Omega_m = 0.254 \pm 0.087, \qquad
\Omega_\Lambda = 0.52 \pm 0.15.
\]
\end{itemize}

Os resultados indicam que a inclusão progressiva de supernovas com menor prob

\end{comment}

In this work we use the supernova sample from DES, in the latest version, DES-Dovekie \cite{popovic2026}. The data comprises 1820 supernovae, including both the SALT3 \cite{kenworthy2021} light curve fit output and the distance modulus estimates from BBC \cite{kessler2017}. 

In ``SALT-like'' light curve fitting, the distance modulus for a supernova is modeled as
\begin{equation}
    \mu_i = m_{Bi} + \alpha x_{1i} - \beta c_i - M + \Delta\mu_{\text{bias,i}},
\end{equation}
where $m_{Bi}$ is peak apparent magnitude in B band, $x_{1i}$ is the stretch, $c_i$ is the color parameter and $\Delta\mu_{\text{bias,i}}$ is the bias correction. Here, we are neglecting the additional parameter $\gamma$, which accounts for the small correlation between brightness and host galaxy mass. $\alpha$, $\beta$ and $M$ can be either fitted simultaneously with the cosmological parameters or estimated in the BBC framework. For this work, we did both approaches.

For all analyses, the dataset was submitted to the following cuts, concerning the light curve fit,
\begin{itemize}
    \item Stretch parameter in the range $|x_1| < 3$;
    \item Error in stretch x1ERR $<1.15$
    \item Color parameter in the range $|c| < 0.3$;
    \item Error in the light curve peak epoch PKMJDERR $ < 2$;
    \item Fit quality FITPROB $ > 10^{-3}$.
\end{itemize}

Folowing \cite{Hlozek2012}, we also tested the results with and without including the theoretical distance modulus in the mean for the Non-Ia distribution, when using the Hlozek model, and tried a second and third order polynomial for the function $\Upsilon$. We also tested the effects of neglecting the bias correction, when fitting the nuisance parameters ($\alpha$, $\beta$ and $M$).

Regarding the type probabilities $P_i$, we tested for the SNIRF \footnote{\url{https://github.com/evevkovacs/ML-SN-Classifier}} and SCONE \cite{Qu2021} estimates, and also investigate the performance when applying cuts for these values, selecting $P_i > 50\%$, $P_i > 90\%$ and all events. Table \ref{modelid} shows the variations implemented in the analyses and the correspondent identifiers used to distinguish them in the remaining tables and figures. We also made tests with the third classifier used by the DES collaboration, SuperNNova (SNN) \cite{moller2020}. However, since the overall performance for all configurations was significantly worse, we decided to exclude it from the final analysis.  

\begin{table}
\begin{center}
\caption{Model identifiers used in the results.} \label{modelid}
\begin{tabular}{p{4.5cm} p{4.5cm} p{4.5cm}}
  \toprule
  Identifier  & Values  & Meaning \\
  \midrule
  Statistical approach &  GMM, Hlozek &  Full Likelihood modeling      \\
  \hline
  Data format & SALT3fit, BBC  &  Fitting ($\alpha$,$\beta$,$M$) or using BBC estimate      \\
  \hline
  Probability estimate & SNIRF, SCONE, MEAN  &  Chosen Ia probability estimator     \\
  \hline
  Probability cut & P90, P50, P0  & Ia probability greater than 90\%, 50\% or 0\%       \\
  \bottomrule
\end{tabular}
\end{center}
\end{table}

Table \ref{ranking_geral} shows the best fit values for the cosmological parameters $\Omega_{m0}$ and $\Omega_{\Lambda 0}$, the values of the logarithmic Bayesian evidence and Bayes factor of each configuration. The reference model for the Bayes factor is the first one. We can see a strong evidence in favor of the our model, for all configurations. It is interesting to note that the configurations using SCONE probabilities are favored for most of the configurations.

\begin{table}[H]
\centering
\footnotesize
\caption{Overall ranking of the tested configurations, according to the Bayesian evidence.}
\begin{tabular}{lcccccc}
\toprule
Model & $\Omega_{m0}$ & $\Omega_{\Lambda 0}$  & $\ln Z$& $\ln \mathcal{B}$ & Jeffreys \\
\midrule

BBC\_GMM\_SCONE\-P50 & $0.316^{+0.050}_{-0.055}$ & $0.798^{+0.104}_{-0.117}$  & 1088.3886 & ----- & ----- \\
BBC\_GMM\_SCONE\_P90 & $0.250^{+0.070}_{-0.079}$ & $0.673^{+0.140}_{-0.157}$  & 1056.3866 & 32.0020 & Very Strong \\
BBC\_GMM\_SNIRF\_P90 & $0.442^{+0.067}_{-0.078}$ & $0.760^{+0.147}_{-0.189}$  & 792.4599 & 295.9287 & Very Strong \\
BBC\_GMM\_SNIRF\_P50 & $0.388^{+0.036}_{-0.045}$ & $0.901^{+0.069}_{-0.107}$  & 708.5352 & 379.8534 & Very Strong \\
SALT3fit\_GMM\_SCONE\_P90 & $0.364^{+0.084}_{-0.103}$ & $0.692^{+0.173}_{-0.227}$  & 632.3504 & 456.0382 & Very Strong \\
SALT3fit\_GMM\_SNIRF\_P90 & $0.512^{+0.077}_{-0.091}$ & $0.668^{+0.180}_{-0.227}$  & 583.1430 & 505.2456 & Very Strong \\
SALT3fit\_GMM\_SCONE\_P50 & $0.413^{+0.080}_{-0.088}$ & $0.592^{+0.170}_{-0.193}$  & 558.8798 & 529.5088 & Very Strong \\
SALT3fit\_GMM\_SCONE\_P0 & $0.324^{+0.047}_{-0.054}$ & $0.791^{+0.105}_{-0.132}$ & 252.6131 & 835.7755 & Very Strong \\
BBC\_GMM\_SNIRF\_P0 & $0.393^{+0.037}_{-0.046}$ & $0.890^{+0.073}_{-0.108}$ & 238.9385 & 849.4501 & Very Strong \\
SALT3fit\_GMM\_SNIRF\_P50 & $0.494^{+0.080}_{-0.094}$ & $0.620^{+0.139}_{-0.239}$  & 175.4539 & 912.9347 & Very Strong \\
SALT3fit\_GMM\_SNIRF\_P0 & $0.523^{+0.080}_{-0.102}$ & $0.642^{+0.196}_{-0.242}$  & -315.5012 & 1403.8898 & Very Strong \\
SALT3fit\_GMM\_SCONE\_P0 & $0.520^{+0.083}_{-0.102}$ & $0.641^{+0.193}_{-0.246}$ & -315.5044 & 1403.8930 & Very Strong \\
BBC\_Hlozek\_SNIRF\_P90 & $0.317^{+0.074}_{-0.095}$ & $0.659^{+0.189}_{-0.231}$  & -525.958 & 1614.3471 & Very Strong \\
BBC\_Hlozek\_SCONE\_P90 & $0.318^{+0.066}_{-0.082}$ & $0.780^{+0.137}_{-0.177}$  & -703.8309 & 1792.2195 & Very Strong \\
BBC\_Hlozek\_SCONE\_P50 & $0.321^{+0.064}_{-0.080}$ & $0.771^{+0.137}_{-0.173}$  & -827.8683 & 1916.2569 & Very Strong \\
BBC\_Hlozek\_SNIRF\_P50 & $0.332^{+0.072}_{-0.085}$ & $0.660^{+0.171}_{-0.201}$  & -1020.1479 & 2108.5365 & Very Strong \\
BBC\_Hlozek\_SCONE\_P0 & $0.325^{+0.061}_{-0.080}$ & $0.793^{+0.128}_{-0.176}$  & -1368.0951 & 2456.4837 & Very Strong \\
BBC\_Hlozek\_SNIRF\_P0 & $0.346^{+0.068}_{-0.082}$ & $0.720^{+0.156}_{-0.204}$  & -1374.7830 & 2463.1716 & Very Strong \\
\bottomrule
\end{tabular}
\label{ranking_geral}
\end{table}

In table \ref{ranking_sem_corte} we compare only the configurations without cuts in the type probabilities. We can see that the configurations with SCONE probabilities are favored when using BBC distance estimates, but not for the SALT3 fit.

\begin{table}[H]
\centering
\footnotesize
\caption{Ranking of the configurations without cut in the type probability, according to the Bayesian evidence.}
\begin{tabular}{lcccccc}
\toprule
Model & $\Omega_{m0}$ & $\Omega_{\Lambda 0}$  & $\ln Z$& $\ln \mathcal{B}$ & Jeffreys \\
\midrule

BBC\_GMM\_SCONE\_P0 & $0.324^{+0.047}_{-0.054}$ & $0.791^{+0.105}_{-0.132}$ & 252.6131 & ----- & ----- \\
BBC\_GMM\_SNIRF\_P0 & $0.393^{+0.037}_{-0.046}$ & $0.890^{+0.073}_{-0.108}$  & 238.9385 & 13.6747 & Very Strong \\
SALT3fit\_GMM\_SNIRF\_P0 & $0.523^{+0.080}_{-0.102}$ & $0.642^{+0.196}_{-0.242}$ & -315.5012 & 568.1143 & Very Strong \\
SALT3fit\_GMM\_SCONE\_P0 & $0.520^{+0.083}_{-0.102}$ & $0.641^{+0.193}_{-0.246}$ & -315.5044 & 568.1175 & Very Strong \\
BBC\_Hlozek\_SCONE\_P0 & $0.325^{+0.061}_{-0.080}$ & $0.793^{+0.128}_{-0.176}$ & -1368.0951 & 1620.7082 & Very Strong \\
BBC\_Hlozek\_SNIRF\_P0 & $0.346^{+0.068}_{-0.082}$ & $0.720^{+0.156}_{-0.204}$  & -1374.7830 & 1627.3961 & Very Strong \\
\bottomrule
\end{tabular}
\label{ranking_sem_corte}
\end{table}

Table \ref{ranking_beams} shows the comparison between the configurations tested for the Hlozek model, using BBC estimates. For these cases, there is strong evidence in favor of SNIRF probabilities, with a 90\% cut in those values. In the same fashion, we made the same comparison for our model and, in this case, there is strong evidence in favor of SCONE probabilities, with a 50\% cut.

\begin{table}[H]
\centering
\footnotesize
\caption{Ranking of the configurations for the Hlozek case, according to the Bayesian evidence.}
\begin{tabular}{lcccccc}
\toprule
Model & $\Omega_{m0}$ & $\Omega_{\Lambda 0}$ & $\ln Z$& $ \ln \mathcal{B}$ & Jeffreys \\
\midrule
BBC\_Hlozek\_SNIRF\_P90&$0.317^{+0.074}_{-0.095}$ &$0.659^{+0.189}_{-0.231}$ &-525.958&-----&-----\\
BBC\_Hlozek\_SCONE\_P90&$0.318^{+0.066}_{-0.082}$ &$0.780^{+0.137}_{-0.177}$ &-703.831&177.872& Very Strong \\
BBC\_Hlozek\_SCONE\_P50&$0.321^{+0.064}_{-0.080}$ &$0.771^{+0.137}_{-0.173}$ &-827.868&301.909&Very Strong \\
BBC\_Hlozek\_SNIRF\_P50&$0.332^{+0.072}_{-0.085}$ &$0.660^{+0.171}_{-0.201}$ &-1020.146&494.189&Very Strong\\
BBC\_Hlozek\_SCONE\_P0&$0.325^{+0.061}_{-0.080}$ &$0.793^{+0.128}_{-0.176}$ &-1368.09503&842.136&Very Strong \\
BBC\_Hlozek\_SNIRF\_P0&$0.346^{+0.068}_{-0.082}$ &$0.720^{+0.156}_{-0.204}$ &-1374.783&848.824&Very Strong \\\bottomrule
\end{tabular}
\label{ranking_beams}
\end{table}

\begin{table}[H]
\centering
\footnotesize
\caption{Ranking of the configurations of the GMM model, when using BBC estimates, according to the Bayesian evidence.}
\begin{tabular}{lcccccc}
\toprule
Modelo & $\Omega_{m0}$ & $\Omega_{\Lambda 0}$  & $\ln Z$& $ \ln \mathcal{B}$ & Jeffreys \\
\midrule
BBC\_GMM\_SCONE\_P50&$0.316^{+0.050}_{-0.055}$ &$0.798^{+0.104}_{-0.117}$ &1088.3886&-----&-----\\
BBC\_GMM\_SCONE\_P90&$0.250^{+0.070}_{-0.079}$ &$0.673^{+0.140}_{-0.157}$ &1056.3866&32.0020&Very Strong \\
BBC\_GMM\_SNIRF\_P90&$0.442^{+0.067}_{-0.078}$ &$0.760^{+0.147}_{-0.189}$ &792.4599&295.9287&Very Strong \\
BBC\_GMM\_SNIRF\_P50&$0.388^{+0.036}_{-0.045}$ &$0.901^{+0.069}_{-0.107}$ &708.5352&379.8534&Very Strong\\
BBC\_GMM\_SCONE\_P0&$0.324^{+0.047}_{-0.054}$ &$0.791^{+0.105}_{-0.109}$ &252.6131&835.7755&Very Strong \\
BBC\_GMM\_SNIRF\_P0&$0.393^{+0.037}_{-0.046}$ &$0.890^{+0.073}_{-0.108}$ &238.9385&849.4501&Very Strong \\\bottomrule
\end{tabular}
\end{table}

We show in table \ref{ranking_salt} the comparison between the configurations performing the fitting for the nuisance parameters $(\alpha, \beta, M)$. In this case, we also find a strong evidence in favor of SCONE probabilities, but with a 50\% cut.

\begin{table}[H]
\centering
\footnotesize
\caption{Ranking of the configurations of the GMM model, when fitting for the nuisance parameters $(\alpha, \beta, M)$, according to the Bayesian evidence.}
\begin{tabular}{lcccccc}
\toprule
Model & $\Omega_{m0}$ & $\Omega_{\Lambda 0}$  & $\ln Z$& $ \ln \mathcal{B}$ & Jeffreys \\
\midrule
SALT3fit\_GMM\_SCONE\_P90&$0.364^{+0.084}_{-0.103}$ &$0.692^{+0.173}_{-0.227}$ &632.3504&-----&-----\\
SALT3fit\_GMM\_SNIRF\_P90&$0.512^{+0.077}_{-0.091}$ &$0.668^{+0.180}_{-0.227}$ &583.1430&49.2073&Very Strong \\
SALT3fit\_GMM\_SCONE\_P50&$0.413^{+0.080}_{-0.088}$ &$0.592^{+0.170}_{-0.193}$ &558.8798&73.4705&Very Strong \\
SALT3fit\_GMM\_SNIRF\_P50&$0.494^{+0.080}_{-0.094}$ &$0.620^{+0.193}_{-0.239}$ &175.4539&456.8964&Very Strong\\
SALT3fit\_GMM\_SNIRF\_P0&$0.523^{+0.080}_{-0.102}$ &$0.642^{+0.196}_{-0.242}$ &-315.5012&947.8516&Very Strong \\
SALT3fit\_GMM\_SCONE\_P0&$0.520^{+0.083}_{-0.102}$ &$0.641^{+0.193}_{-0.246}$ &-315.5044&947.8548&Very Strong \\
\bottomrule
\end{tabular}
\label{ranking_salt}
\end{table}

\begin{figure}[H]
    \centering
    \includegraphics[width=0.45\textwidth]{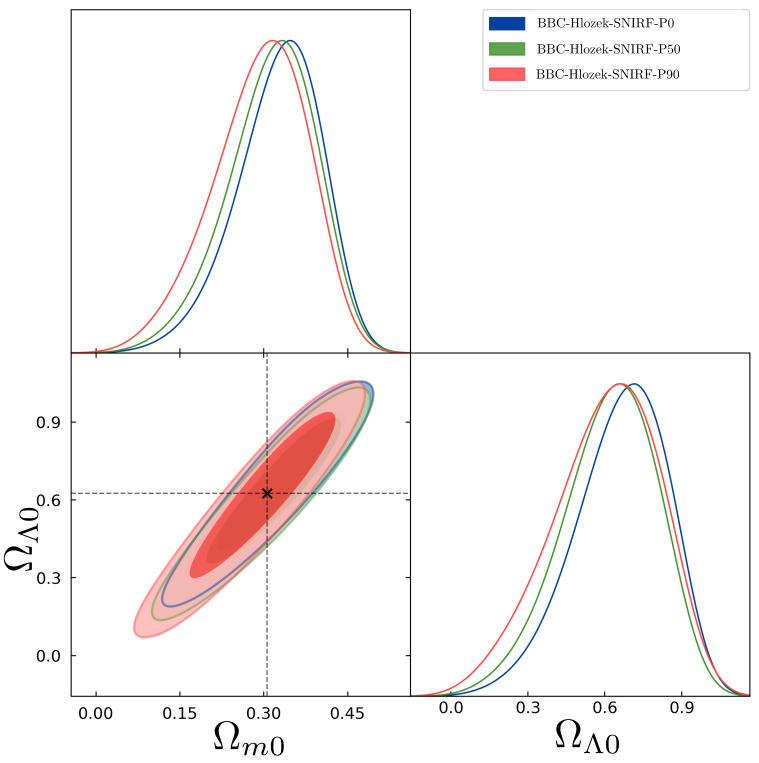}
    \includegraphics[width=0.45\textwidth]{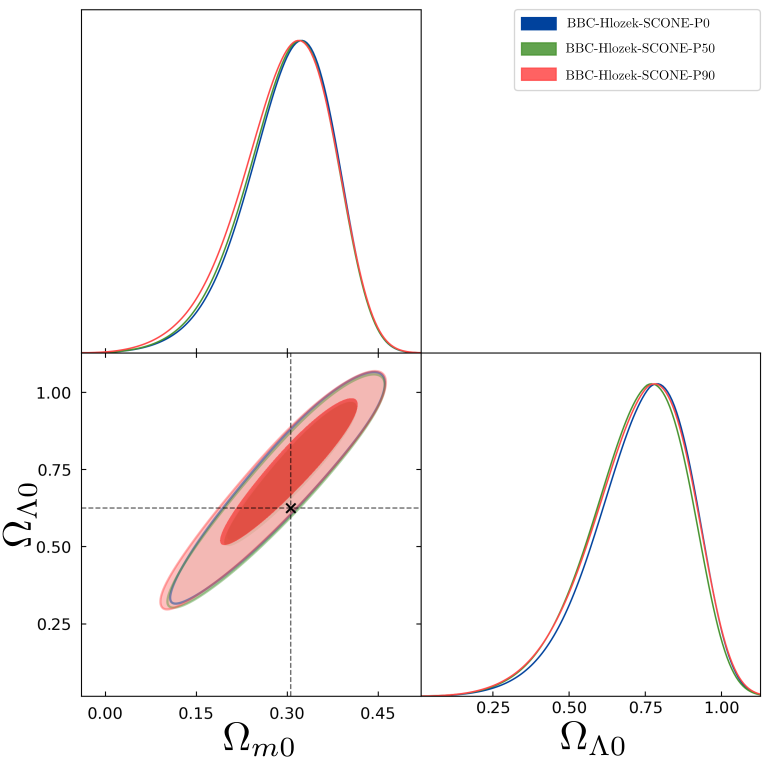}
    \caption{The figure shows 68\% and 95\% confidence contours and the marginalized distributions of the cosmological parameters for the Hlozek model, when using SNIRF (left panel) and SCONE (right panel) probabilities, considering three configurations: without cut the type probability (blue), with $P > 50\%$ (green) and with $P > 90\%$ (red). All results are marginalized over all other free parameters. The black cross is the best fit for the Pantheon+ \cite{Brout2022}, for visual reference.}
    \label{comparacao_BEAMS_0_50_90}
\end{figure}

In figure \ref{comparacao_BEAMS_0_50_90} we show the confidence contours and marginalized 1-D distributions for the cosmological parameters, considering all configurations for Hlozek model. In the left panel we compare different cuts for SNIRF probabilities while in the right panel we show the same comparison for SCONE ones. We can excellent agreement between the results for different cuts, but there is a small shift between the contours for different classifiers, although they agree at 68\% of confidence level.

\begin{figure}[H]
    \centering
    \includegraphics[width=0.45\textwidth]{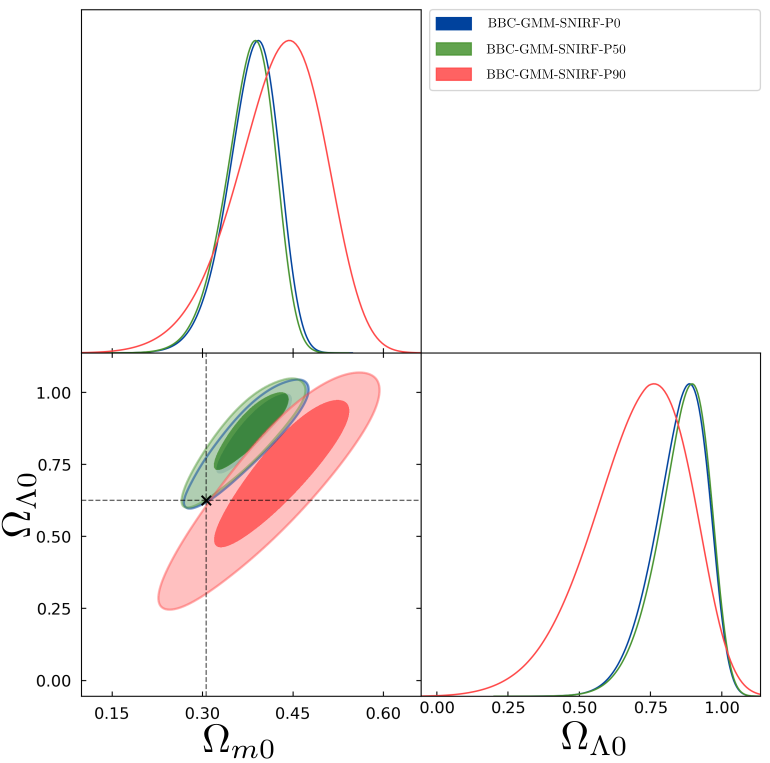}
    \includegraphics[width=0.45\textwidth]{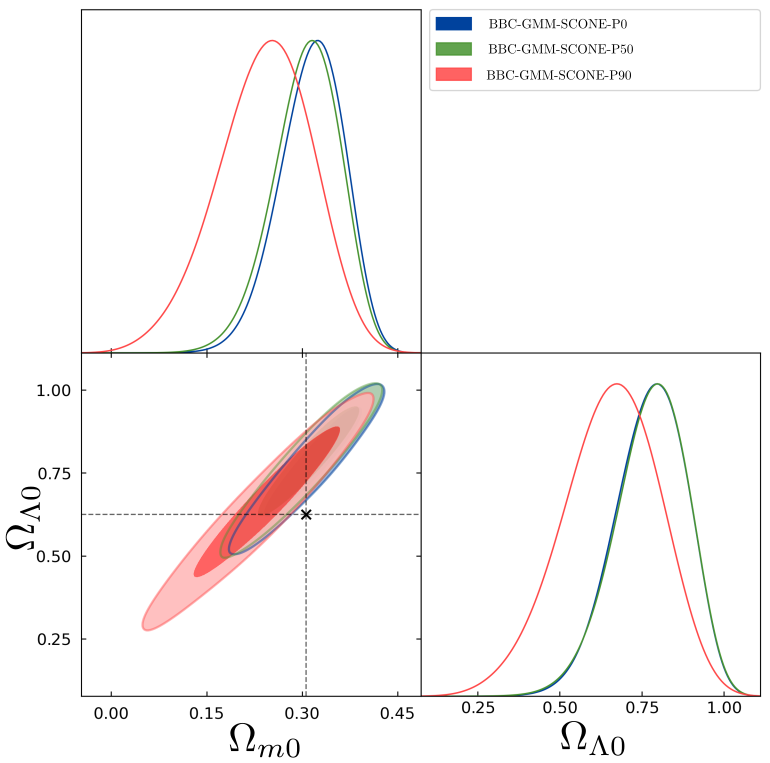}
    \caption{The figure shows 68\% and 95\% confidence contours and the marginalized distributions of the cosmological parameters for GMM model, when using SNIRF (left panel) and SCONE (right panel) probabilities, considering three configurations: without cut the type probability (blue), with $P > 50\%$ (green) and with $P > 90\%$ (red). All results are marginalized over all other free parameters. The black cross is the best fit for the Pantheon+ \cite{Brout2022}, for visual reference.}
    \label{comparacao_gmm_0_50_90}
\end{figure}

\begin{figure}[H]
    \centering
    \includegraphics[width=0.45\textwidth]{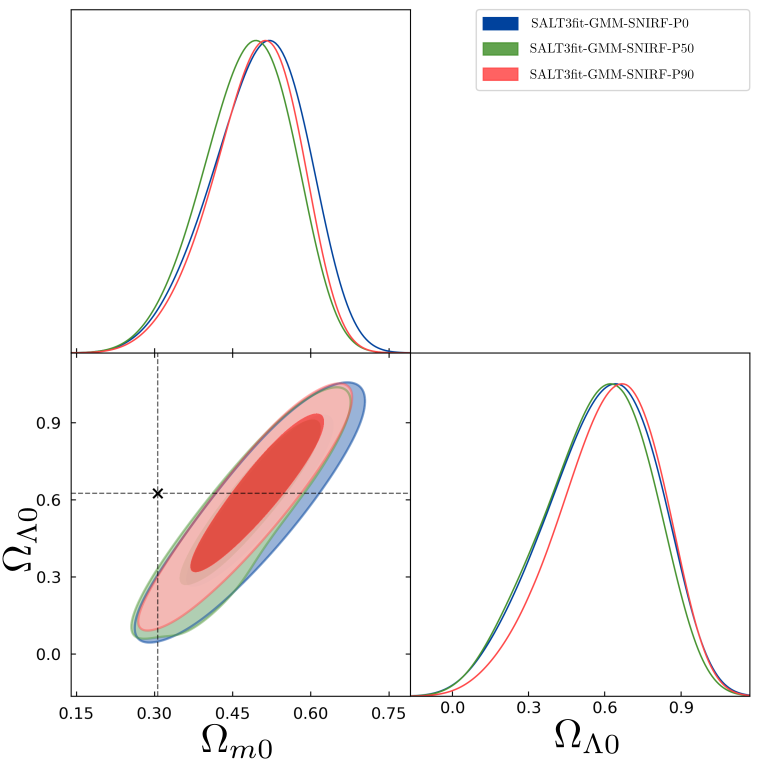}
    \includegraphics[width=0.45\textwidth]{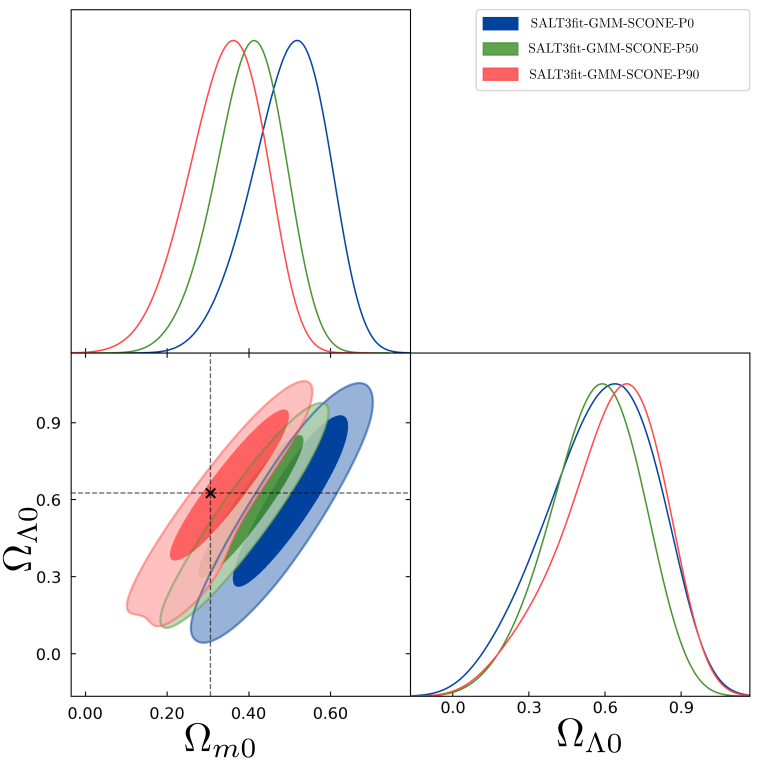}
    \caption{The figure shows 68\% and 95\% confidence contours and the marginalized distributions of the cosmological parameters for GMM model, when using SNIRF (left panel) and SCONE (right panel) probabilities, considering three configurations: without cut the type probability (blue), with $P > 50\%$ (green) and with $P > 90\%$ (red). In this case, we performed a simultaneous fit with the SALT3 nuisance parameters ($\alpha$, $\beta$, $M$). All results are marginalized over all other free parameters. The black cross is the best fit for the Pantheon+ \cite{Brout2022}, for visual reference.}
    \label{comparacao_salt3_0_50_90}
\end{figure}

In figure \ref{comparacao_gmm_0_50_90} we show the confidence contours and marginalized 1-D distributions for the cosmological parameters, considering all configurations for GMM. In this case, we can see a curious shift between the result for 90\% cut and the others, when using SNIRF probabilities. Such distinction does not appear for the configurations with SCONE. In figure \ref{comparacao_salt3_0_50_90} we show that GMM is still more sensitive to probabilities cuts, but we now observe shifts between cuts for SCONE and not for SNIRF.

\begin{figure}[H]
    \centering
    \includegraphics[width=0.45\textwidth]{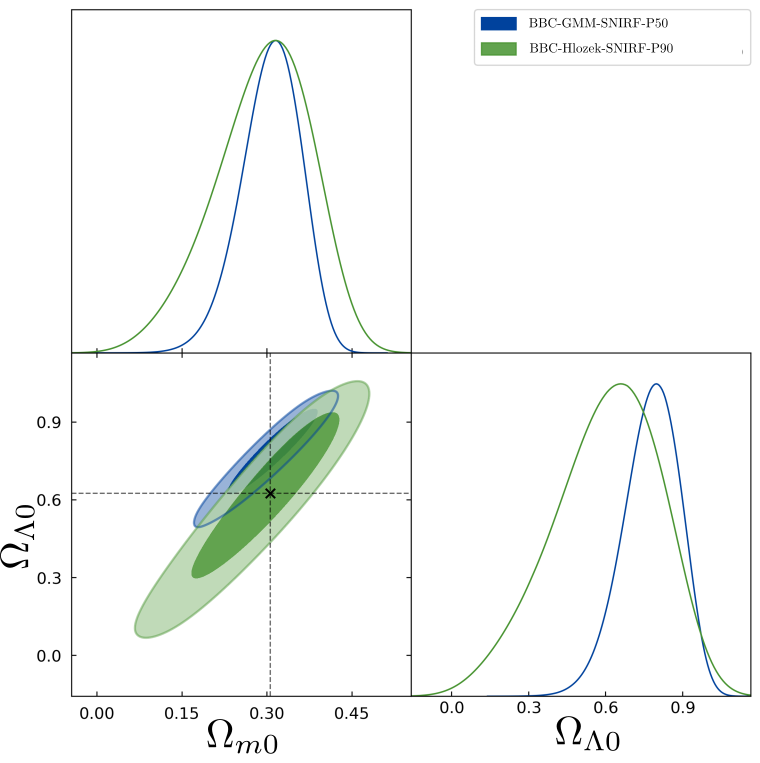}
    \includegraphics[width=0.45\textwidth]{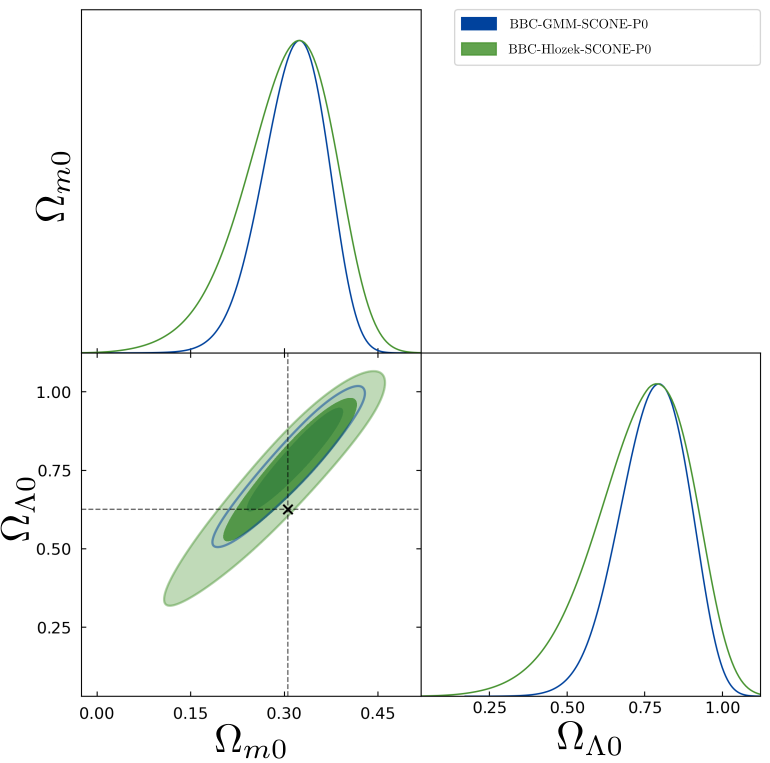}
    \caption{The figure shows 68\% and 95\% confidence contours and the marginalized distributions of the cosmological parameters for the best configuration for GMM (blue) and Hlozek (green) models, when using SNIRF (left panel) and SCONE (right panel) probabilities. All results are marginalized over all other free parameters. The black cross is the best fit for the Pantheon+ \cite{Brout2022}, for visual reference.}.
    \label{melhor_resultado_GMM50_BEAMS90}
\end{figure}

In figure \ref{melhor_resultado_GMM50_BEAMS90} we compare the best configuration for GMM and Hlozek models. In the left panel, we show the results for SNIRF probabilities and, in the right panel, for SCONE. Although they are in agreement in both cases, we can see a slight displacement, in SNIRF case, towards higher values of $\Omega_{\Lambda 0}$ for our model. However, the agreement for SCONE case is remarkable.

Based on the changes in the results when we use different classifiers, we decided to test another configuration, in which we use as probability the mean of the estimates given all classifiers. For this analysis, we decided to include SNN. We observed that, for some events, the classifiers can strongly disagree. To ensure a certain level of agreement among the classifiers, we impose a consistency criterium based on the standard deviation of these estimates. To this end, we selected events for which all probabilities obey the condition
\begin{equation}
    |P_i - P_{mean}| \leq \sigma_P.
    \label{mean_cut}
\end{equation}
For this analysis we refrained from applying any additional cut in the probabilities. 

\begin{figure}[H]
    \centering
    \includegraphics[width=0.8\textwidth]{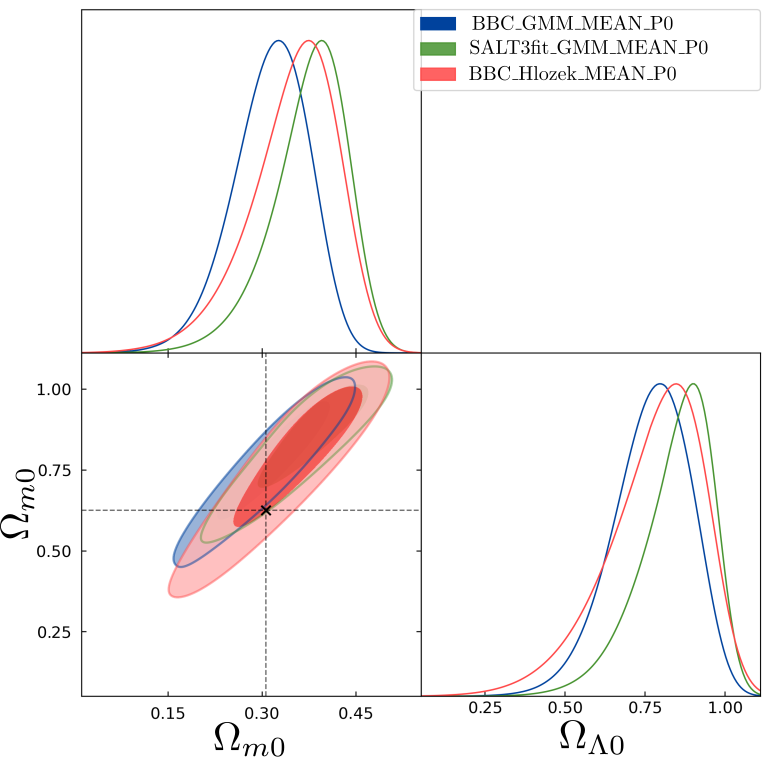}
    \caption{The figure shows 68\% and 95\% confidence contours and the marginalized distributions of the cosmological parameters for GMM, using BBC (blue) and fitting the nuisance (SALT3) parameters (green), and Hlozek (red) models, when using the mean of probabilities from SCONE, SNIRF and SuperNNova. All results are marginalized over all other free parameters. The black cross is the best fit for the Pantheon+ \cite{Brout2022}, for visual reference.}.
    \label{grafico_media_P}
\end{figure}

Table \ref{ranking_mean} show the classification for the three configurations tested, Hlozek with BBC estimates, GMM with BBC estimates and GMM with SALT3 fit. Figure \ref{grafico_media_P} shows the confidence contours and marginalized distributions for the cosmological parameters, for these same cases.
The most notable feature is that the discrepancies between the approaches pointed out in the previous results are considerably alleviated in this case, with the GMM model exhibiting more constraining power for the same dataset.

\begin{table}
\caption{Ranking of the configurations tested using the mean type probability, Hlozek with BBC estimates, GMM with BBC estimates and GMM with SALT3 fit.}
\centering
\begin{tabular}{llrrl}
\toprule
model & $\ln Z$ & $\ln \mathcal{B}$ & Jeffreys \\
\midrule
BBC\_GMM\_MEAN\_P0  & 988.2431 & --- & --- \\
SALT3fit\_GMM\_MEAN\_P0  & 818.1877 & 170.0554 & Forte \\
BBC\_Hlozek\_MEAN\_P0  & -743.0351 & 1731.2782 & Forte \\
\bottomrule
\end{tabular}
\label{ranking_mean}
\end{table}

\section{Conclusion}
In this work we investigated the impact of the statistical modeling of the contamination in a photometric supernova dataset on the cosmological constraints. We proposed a simplified likelihood based on the assumption that, if the contamination is a small fraction of the sample, its primary neat effect may be a redshift dependent change in the mean of the distribution. We tested this hypothesis against the usual two-component model, under the same conditions and found the new approach is favored for all tested configurations. The current results indicate that the proposed model is a promising way of improving the constraining power of photometrically classified supernova data.

\acknowledgments
The authors thank Dr. C\'{a}ssia da Silva Nascimento for the fruitful discussions and support. RRRR thanks CNPq for partial financial support (grant no. $303829/2025-7$). 
This project used public archival data from the Dark Energy Survey (DES) \cite{descollaboration2025,Sanchez_2024,Vincenzi_2024,popovic2026}, available at \url{https://github.com/des-science/DES-SN5YR}.
The analysis presented in this work was implemented in Python, taking advantage of the following codes: SciPy \cite{scipy}, emcee \cite{emcee}, matplotlib \cite{Hunter:2007}, getdist \cite{Lewis_2025}.

%\newpage 
%\printbibliography
% Bibliography

%% [A] Recommended: using JHEP.bst file
\bibliographystyle{JHEP}
\bibliography{Bibliography.bib}

\end{document}